\documentclass[a4paper]{article}

\usepackage{18lomcon} 
\usepackage{cite}  
\usepackage{epsfig}  
\usepackage{epstopdf}

\begin{document}


\title{KINEMATICS OF SPIRALS AS A PORTAL TO THE NATURE OF DARK MATTER  }

\author{Paolo Salucci \email{salucci@sissa.it}
}

\affiliation{SISSA, Via Bonomea 265, Trieste, Italy}


\date{}
\maketitle


\begin{abstract}
The gravitational field of Spiral galaxies is well traced by their rotation curves. Only recently it has become 
of extreme interest that the latter form a family ruled by two parameters of 
the {\it luminous} component: the disk
length-scale $R_D$ and the magnitude $M_I$.
This evidence is so strong and consequential that it must be taken as the starting point for the 
investigation on the issue of the dark matter in galaxies. The emerging fact is that structural quantities deeply rooted in the
luminous components, like the disk lenghtscales are found to tightly correlate with structural quantities 
deeply rooted in the dark component, like the DM halo core radii. These unexpected evidences may strongly call for a shift
of
paradigm with respect to the current Cold collisionless Dark Matter one. 
\end{abstract}

 \section{Introduction} 

 Recently, a number of discoveries seems to have weakened our certainties about the dark matter,
the elusive substance believed to be a particle, that constitutes about $25 \% $ of the mass energy of the Universe, 
plays a crucial role in the formation of structures, and, since late 70's, is thought to surround the
luminous disks of Spirals (\cite{rubin}), interacting with the ordinary matter only by gravitation. In disk systems, the 
circular velocity balances the radial variations of the total gravitational potential $\Phi$: 
 $V^2= r \ d\Phi/dr$, while the Poisson Equation relates the galaxy gravitational potential $\Phi= \sum_i \Phi_i$ and the
density of the mass components
$\nabla^2 \Phi_i= 4 \pi G \rho_i$, 
where $\rho_i $ stands for dark halo, stellar disk, stellar bulge, HI disk surface/volume densities (more specifically, $\rho_h(r),
\rho_{bu}(r), \mu_{d}(R), \mu_{HI}(R) $) with $R,z$ the cylindrical
coordinates.

The main component of the luminous matter in spirals is the well-known Freeman 
stellar disk of exponential surface density: \cite{freeman}
\begin{equation}
\mu_D (r) = \frac {M_D} {2 \pi R_D} e^{-r/R_D}
\end{equation}
 with $R_D$ the disk length-scale and $M_D$ the disk mass). 
\footnote {It is useful to define $R_{opt} \equiv 3.2 R_D$ as the optical size of the stellar disk, as the radius encompassing 83 $\%$ of the total galaxy
luminosity. }
We have within r.m.s. of 0.1 dex (\cite{Tonini}):
$
\log\left(\frac{R_D}{kpc}\right) = 0.633+0.379\log\left(\frac{M_D}{10^{11}M_{\odot}}\right) 
+ 0.069\left(\log \frac{M_D}{10^{11}M_{\odot}} \right)^2,
$
that links the stellar disk masses $M_D$ with their length-scales $R_D$.
 The HI disk has a surface density profile like Eq(1), with a length
scale $R_{HI}=3\ R_{D}$ \footnote {The $M_{HI}$ vs $M_D$ relationship is in \cite{e}}. However, since the HI component is
negligible  
inside $R_{opt}$, it not will be further considered. 

\begin{figure}
\vskip -5.3cm
 \begin{center}
\epsfig{file=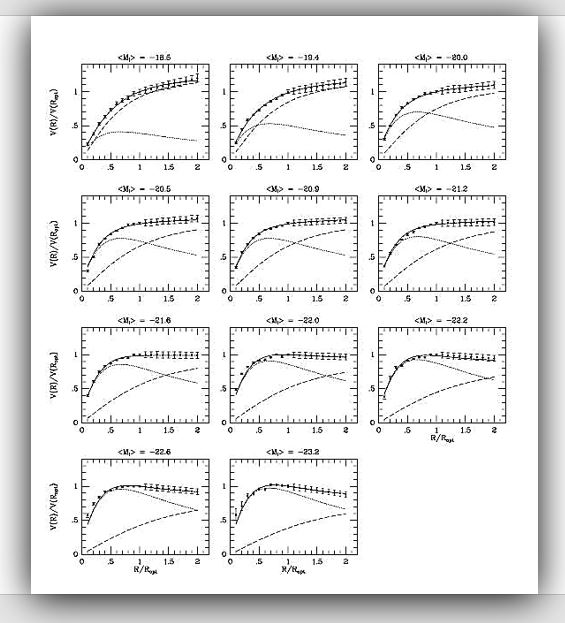, width=7.0 cm}
\vskip 0.9cm
\epsfig{file=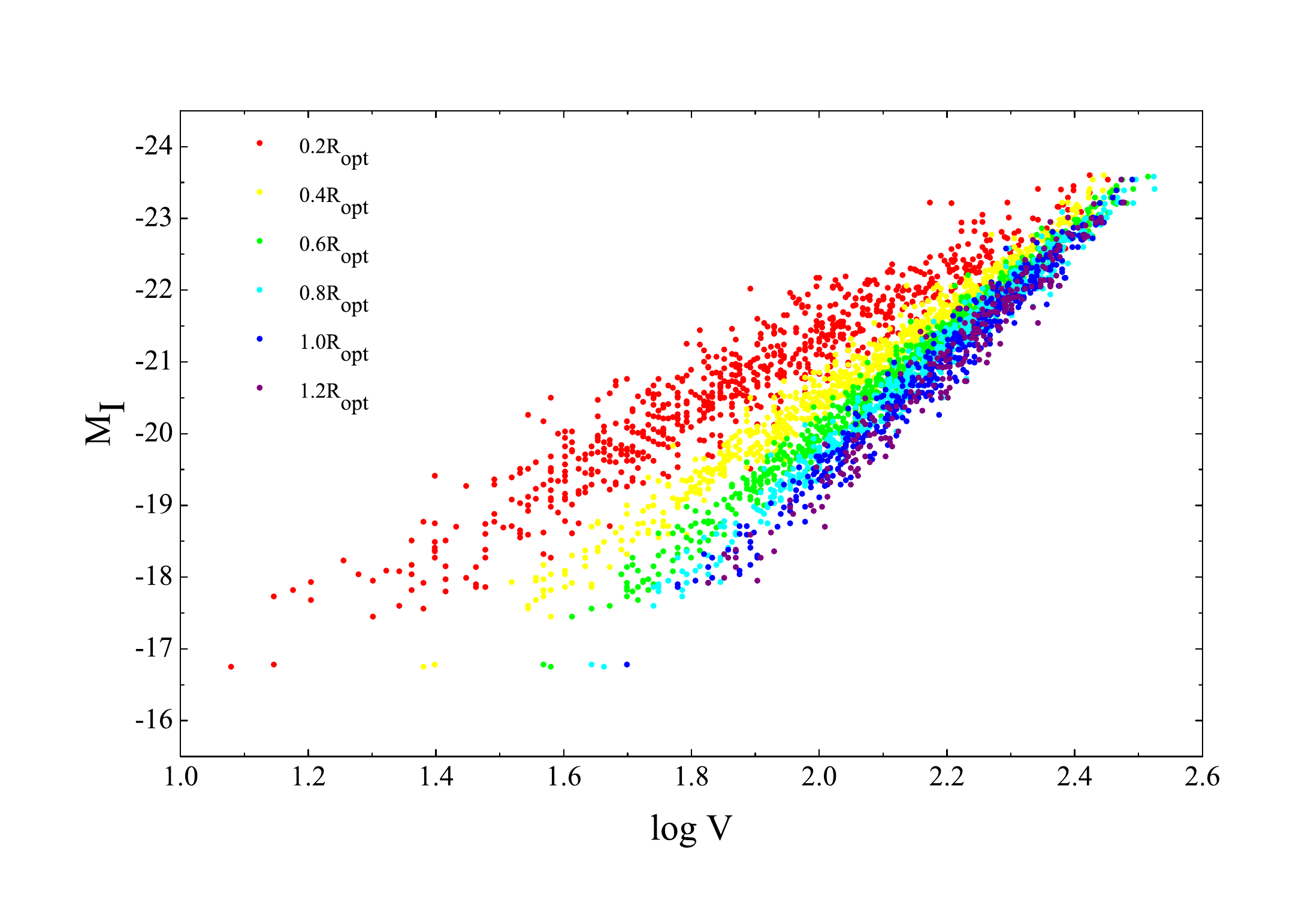, width=10.2 cm}
\end{center}
\caption{({\it up}) The URC ({\it black lines}) and the coadded RCs ({\it points}).({\it bottom}) The Radial TF}
\end{figure}

\section {The Universal Rotation Curve from the coadded rotation curves of 967 Spirals}

We can represent the rotation curves of Spirals by means of the Universal Rotation Curve (URC) pioneered in
\cite{rubin,p91} and set in \cite{pss,s7}. 
By adopting the normalized radial coordinate $x\equiv r/R_{opt}$ the RCs of Spirals are well described by a Universal profile,
function of $x$ and of $\lambda$, where $\lambda $ is one among $M_I$,
the I magnitude, $M_D$, the disk mass and $M_{vir}$, the halo virial mass (\cite {s7}).
 
An universal magnitude-dependent profile is evident in the 11 {\it coadded} rotation curves $V_{coadd}(x, M_I)$(see Fig (1) and \cite{pss}) built from the individual RCs of a sample of 967 Spirals. I-band surface photometry measurements \cite{PS95}
provide these objects with their stellar disk length scales $R_D$. 
The URC is built in a three steps way {\bf 1)} The I magnitude range of Spirals is divided in 11 successive bins centred at
$M_I$ as listed in 
Table 1 of \cite{pss}. {\bf 2)}The RC of each galaxy is assigned to its corresponding I magnitude bin, normalized by its $V(R_{opt})$ value and then expressed in terms of its normalized radial coordinate 
$x $. {\bf 3.)} The double-normalized RCs $V(x)/V_{opt}$ curves are coadded in 11 magnitude bins and
in 20 radial bins of length 0.1, and then averaged to get: $V_{coadd}(x, M_I))/ V_{coadd}(1, M_I)$, see Fig (1). The 11
values of $ V_{coadd}(1, M_I)$ are given in Table 1 of \cite{pss}.

\begin{figure}[t]
\centering
\includegraphics[height=8.2cm]{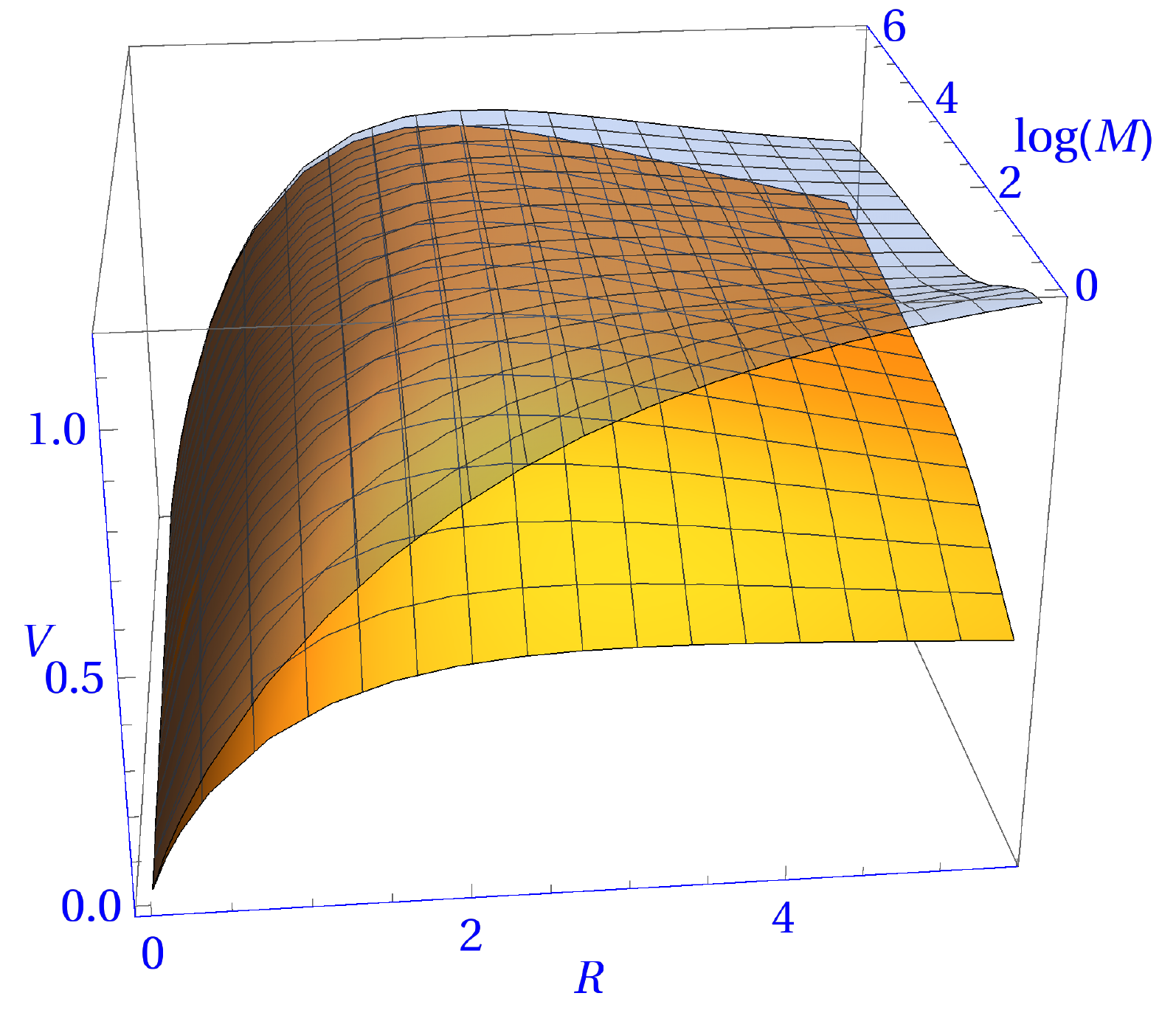}
\caption{The URC obtained from the coadded RCs {\it blue}. $V$ is in units of $V_{opt}$, $R$ in units of $R_D$ and the virial halo mass $M$ in units of $10^{11} M_\odot $. Also shown the baryonic component ({\it orange})
Auxiliary to Eq(5) we have: $log \ (M_D/M_\odot) = -0.52
M_I-0.45$ and $log \ (M_{vir}/M_\odot)= 25.53 - 10/3 \ Log\ M_D+ 1/5 \ (log\ M_D)^2$ that connects $M_I,M_D, M_{vir}$}
\end{figure}%

The Universal Rotation Curve $V_{URC}(x, M_I) $ is the (halo+disk) {\it physical} velocity model that
very well fits the above $V_{coadd}(x, M_I)$. At any $x$ and for any $M_I$: $ V^2_{URC}=
V^2_{URCd}+ V^2_{URCh}$. From Eq(1) the disk velocity component is: 
\begin{equation}
 V_{URCd}^2(x)=\frac{ G M_D}{2 \ R_{D} } \left(3.2x\right) ^{2}\left(I_{0}K_{0}-I_{1}K_{1}\right) 
\end{equation}
with the Bessel functions evaluated at $1.6 \ x$. For the DM halo we assume the Burkert density profile \cite{SB} : 
\begin{equation}
 \rho_{URCh}(r)=\frac{\rho_0 r_0^3}{(r+r_0)(r^2+r_0^2)}
\end{equation}.

where $r_0$ the core radius and $\rho_0$ the central halo density are its free parameters. The velocity profile reads \footnote{In this framework, the galaxy virial radius $R_{vir} $ is defined as the solution of the equation: $G^{-1} \ V^2_{URCh} (R_{vir}) R_{vir} = 100 \times 4/3 \pi \ 100 \ R_{vir}^3$ 
where $\rho_c=1.8 \ 10^{-29}g/cm^3$ is the critical density of the Universe. The virial mass is then: $ M_{vir} = 10^{12} M_\odot (R_{vir}/260 \ kpc)^3$.}
:

\begin{equation}
V^2_{URCh}(r) =6.4\frac{\rho_0 r_0^3}{r}(\ln (1+\frac{r}{r_0})-\arctan \frac{r}{r_0}) +\frac{1}{2}\ln (1+\frac{r^2}
{r_0^2})). 
\end{equation}

In detail, $V_{URC}(x, \rho_0,r_0,M_D) = (V_{URCd} (x;M_D)^2+ V_{URCh} (x;\rho_0;
r_0)^2)^{1/2} $ fits very well the coadded RC, see Fig(1). The free parameters result all related to each other
\cite{s7} (masses in $M_\odot$, densities in g/cm$^3$, distances in kpc):

$$
log {\rho_0\over \mathrm{g}~\mathrm{cm}^{-3}} = -23.5-0.96 \left({M_D}\over 10^{11}\, \rm
M_{\odot}\right)^{0.31} 
$$
\begin{equation}
log \left(\frac{r_0}{\mbox{kpc}}\right) \simeq 0.66 +0.58 \log\left(\frac{M_{vir}}{10^{11}M_{\odot}}\right)
\end{equation}

\section{The Universal Rotation Curve from the Radial Tully Fisher relationship in three large samples of Spirals}
 
Yegorova and Salucci (2007), by analysing three samples of 794, 86 and 91 spirals of various Hubble types, 
discovered what was named as the Radial Tully-Fisher (RTF)
relation (see \cite{ys} and Fig (1)). This is an ensemble of tight relationships between the galaxy absolute magnitude $M_I$
and
$\log V_n$, the logarithm of circular velocity measured, in each object, at the same fixed fraction of the optical radius
$R_n/R_{opt}$ ($R_{opt} \equiv 3.2 R_D$). In detail, from \cite{ys} we have: 
\begin{equation}
V_n\equiv V(R_n) , \ \ \ \ R_n= (n /10) \ R_{opt}, \ \ \ \ 1\leq n \leq 10, \ \ \ \ \ 
M_{I} = a_n \log V_n + b_n 
\end{equation}

\begin{figure}[t]
\begin{center}
\includegraphics[height=8.7cm]{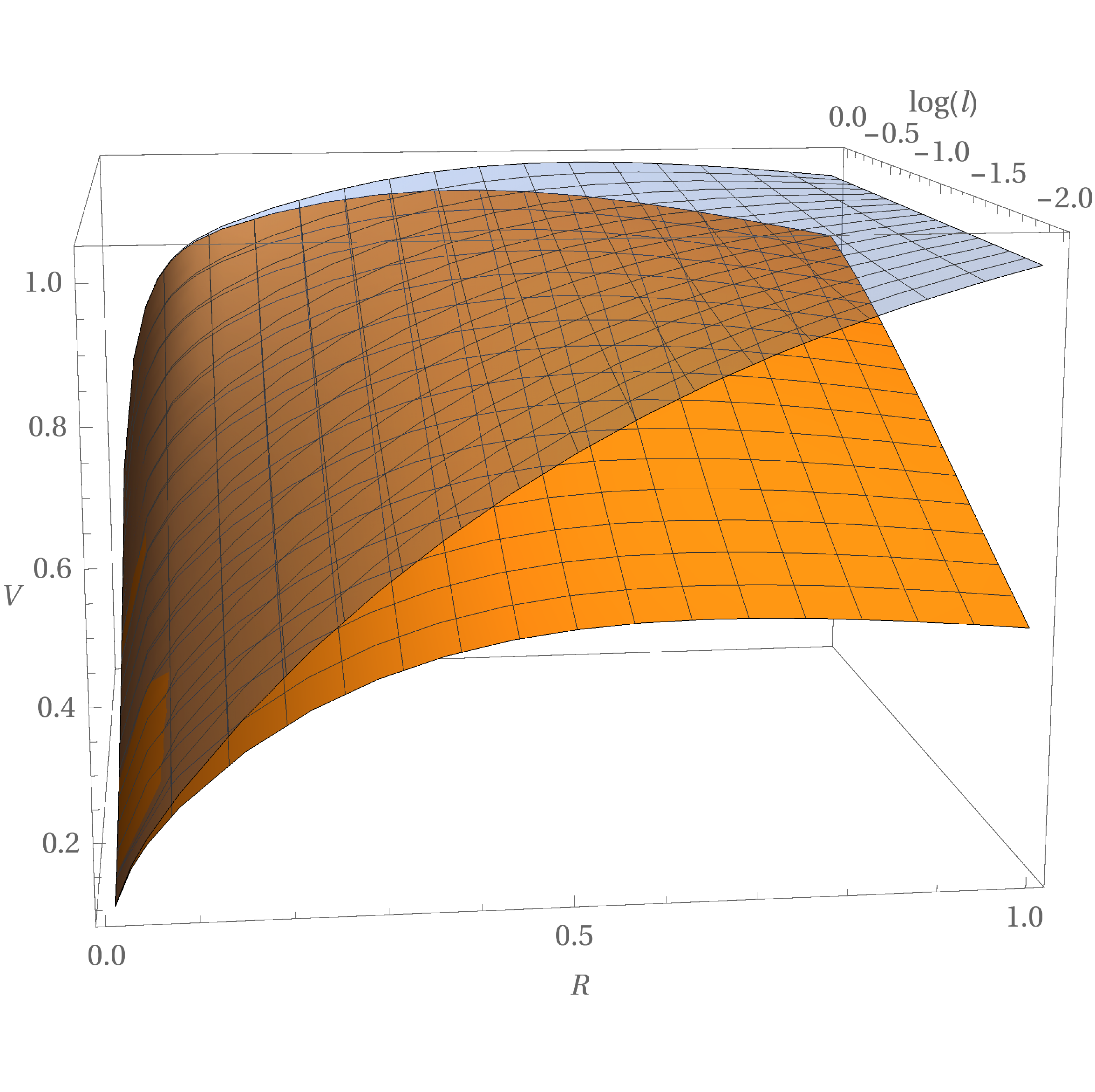}
\end{center}

\caption{The URC from the Radial Tully Fisher ({\it blue}). $V$ is expressed in units of $V_{opt}$, $R$
in units of $R_{opt}$. Also shown the baryonic component ({\it orange}). To compare it with the URC in Fig (2), 
one uses: $\rho_0=1/(4 \pi G) (V_{opt}/R_{opt})^2
\frac{1+\alpha}{\alpha}$ and $r_0= 0.8\ \alpha \ R_{opt}$}

\end{figure}

where $n$ indicates the position of the center of the radial velocity bin and $a_n$,$b_n$ the values
of the parameters of the $n$th fit. All relationships of Eq. (10) result statistically relevant
(see Eq(1) and \cite{ys}) and, remarkably, in all three samples, $a_n$ shows the same strong radial variation with radius: \cite{ys} 
\begin{equation}
a_n=-2.3 - n + 0.04 \ n^2 
 \end{equation}
 with a small statistical uncertainty. Eq (2) implies that, in spirals, light does
not trace the distribution of the gravitating matter which would require: $a_n\simeq -7.5$. 
Eq(6)-(7), instead, imply the existence of a Universal velocity model, function of $x$ and $M_I$. The URC we build from of Eq(6) has 3 components (disk, bulge, halo). At any $x$
we have: $V^2_{URC}= V^2_{URCd}+ V^2_{URCh}+
V^2_{URCbu}$. In the following, without any loss of generality, we assume (\cite{pss,ys}): $ R_D =R_1 l^{0.5}$ and $M_D =M_1 l^{1.3}$. From Eq(8) we have: \footnote{$l\equiv 10^{(M_I-M_I^{max})/2.5}$
with $M_I^{max}=-23.5 $. The constants $R_1=5.6$ \
kpc, $M_1=2 \times 10^{11} M_\odot$ and $ V_1^2
\equiv G \ M_1/R_1=500^2 (Km/s)^2$ play no role in the evaluation of the $a_n$'s.} 
 $V_{URCd}^2(x,l) = V_1^2 l^{0.8} f_d(x)$
 with $ f_d(x)={1\over {2}}(3.2 x)^2(I_0(1.6 x)K_0(1.6 x)-I_1(1.6 x)K_1(1.6 x))$ and: $V^2_{d}(1,1)= 0.347 \ V_1^2 $.

For the bulge we set that, at $R_{opt}$, $V^2_{URCbu} $ is $c_{bu}/(3.2\cdot 0.347)\ l^{0.5}$
times the disk contribution, $c_{bu}$ is a free parameter and the exponent $0.5$ is suggested
by the bulge-to disk vs total luminosity relation in spirals. Since all kinematical data refer to
radii outside the bulge half light radius, we can consider this component as a point mass. Then: $ V_{RTFbu}^2(x,l) = \ c_{bu} \ V_{RFTd}^2(1,l) \ l^{0.5} x^{-1}$
\footnote{No result changes by adopting a more realistic Sersic profile}.

For the halo velocity contribution we adopt: $V_{URCh}^2(x) = V_1^2 \ f_h(x;\alpha) A(l)$,
 where $\alpha$ is the halo velocity core radius in units of $R_{opt}$, and: $f_h(x;\alpha) = ({x^2\over {x^2+\alpha^2}})(1+\alpha^2)$. This is the simplest profile that describes the DM halo density distribution inside $R_{opt}$. Notice that for values of 
$\alpha \simeq 1.2$ and $\alpha<1/3$, $f_h$ reproduces the Burkert and the NFW velocity profiles. We assume a power law
between
halo mass and luminosity: $M_h(1,l) \propto l^{k_h} $, where 
$k_h $ is a free parameter. We normalize the latter by setting that, at $x=1$, $
V_{URCh}^2(1,l)$ is $ c_h/(3.2\ 0.347) l^{(k_h - 0.5)} $ times the disk
contribution $ V_{URCd}^2(1,l)$. The quantity $(0.9 \ c_h)^{1/2}$ is then the halo-to-disk fraction at $(l,x)= (1,1) $ and $A(l)=c_h l^{k_h-0.5}$. Then, the URC, aimed to best fit Eq(7), is written as:
 \begin{equation} 
V_{URC}^2 (x,l ; \alpha, c_{bu}, c_h, k_h)= V_1^2 (c_{bu} l^{1.3}/x + l^{0.8} f_d(x) + A(l) f_h(x;\alpha))
\end{equation}

The best fit values of the parameters $\alpha$, $c_{bu}$, $c_h$ and $ k_h $, that specify completely the URC are (\cite{ys}):
$k_h=0.79 \pm 0.04$, $c_{bu}=0.13 \pm 0.03$, $c_h=0.13 \pm 0.06$ and $\alpha=1 ^{+1}_{-0.5}$. They imply that, inside $R_{opt}$, less luminous galaxies have a larger fraction of dark matter: $M_D/M_h(1,l)\propto
l^{0.5}$ and point towards {\it cored} DM halos.

\section{Discussion and Conclusions} 

We realize that, once expressed in terms of the same variables, the URCs of Fig(2) and Fig (3), 
built in very different ways, almost coincide. The only difference is in the very inner regions of highest luminosity objects for which the
latter URC has an additional contribution from a stellar bulge. In the 3D space, whose axes are $V, x, M_I$: {\it i)} at
any magnitude $M_I$, the coadded RC $V_{coadd}(x, M_I)$ and {\it ii)} at any fixed normalized radius $x$, 
the Radial Tully Fisher relationship: $M_I = a_x +b_x log (V(x))$, are the normal sections of a surface $V(x,M_I)$, whose
analytical representation, in terms of dark and luminous velocity components, is what we call the Universal Rotation
Curve. 

\begin{figure}[t]
\begin{center}
\includegraphics[height=6.4cm]{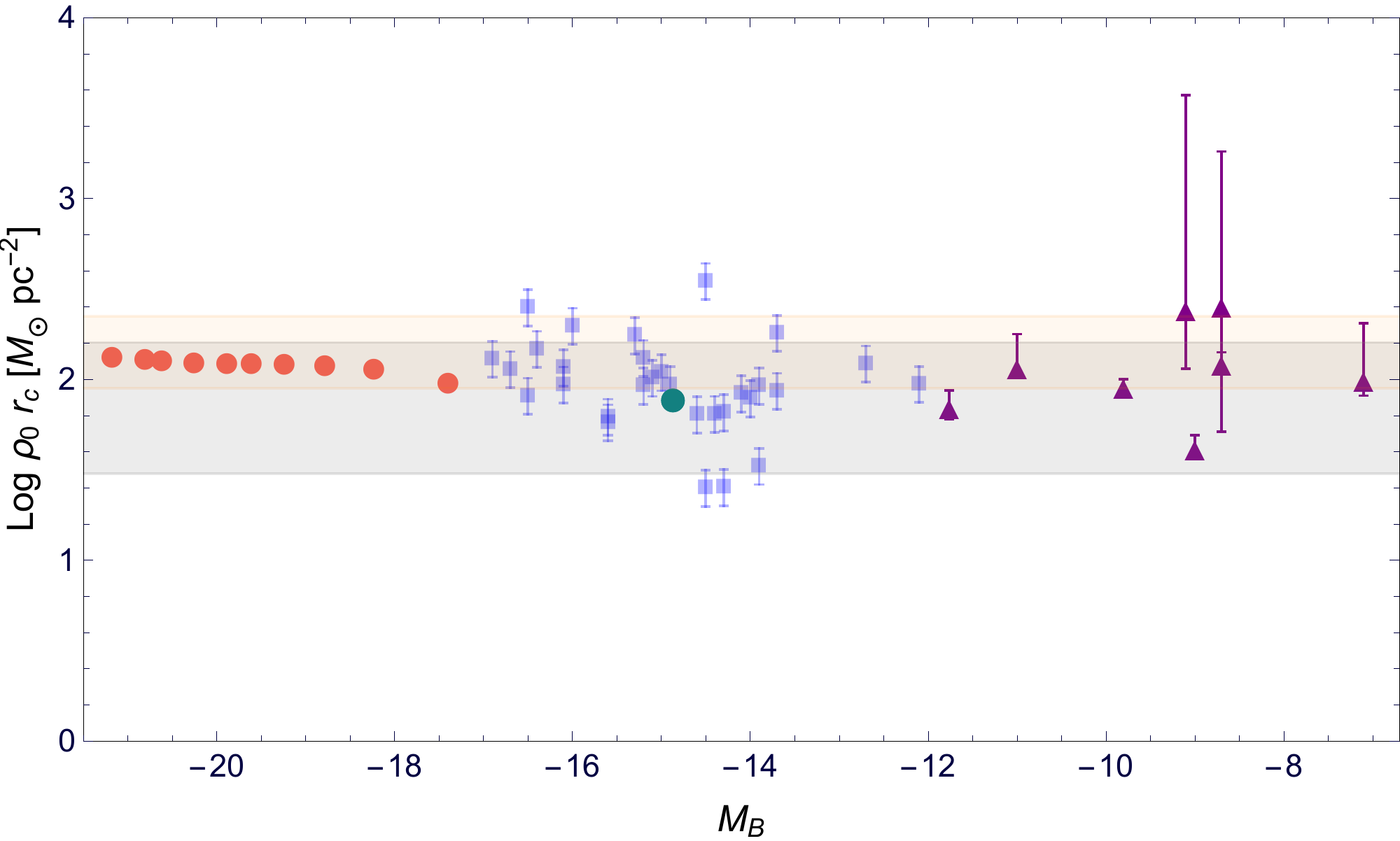}
\end{center}
\caption{ The halo central surface density $\rho_0 r_c$ for different samples of Spirals}
\end{figure}%

This result, obtained by means of two independent lines of investigation of the kinematics of thousands of disk systems, has very important consequences: 

$\bullet$ The non-dimensional radial coordinate $x$ gauges and links together all spirals: 
but this occurs independently on whether, at a radius $x$, $V(x)$ is dominated by the dark or by the luminous matter
component.
 
$\bullet$ All spirals show a dark halo component with a constant density inner core of size $r_0$ 
that surrounds a stellar thin disk of lenghtscale $R_D$. If this would not be enough, the two lenghtscales result closely
correlated see \cite{D}. 

$\bullet$ The central halo density $\rho_0$ and the core radius $r_0$ emerge very close correlated, so that their product is constant in galaxies of very different luminosity. 

\begin{figure}
\begin{center}
  \includegraphics[width=0.96\textwidth]{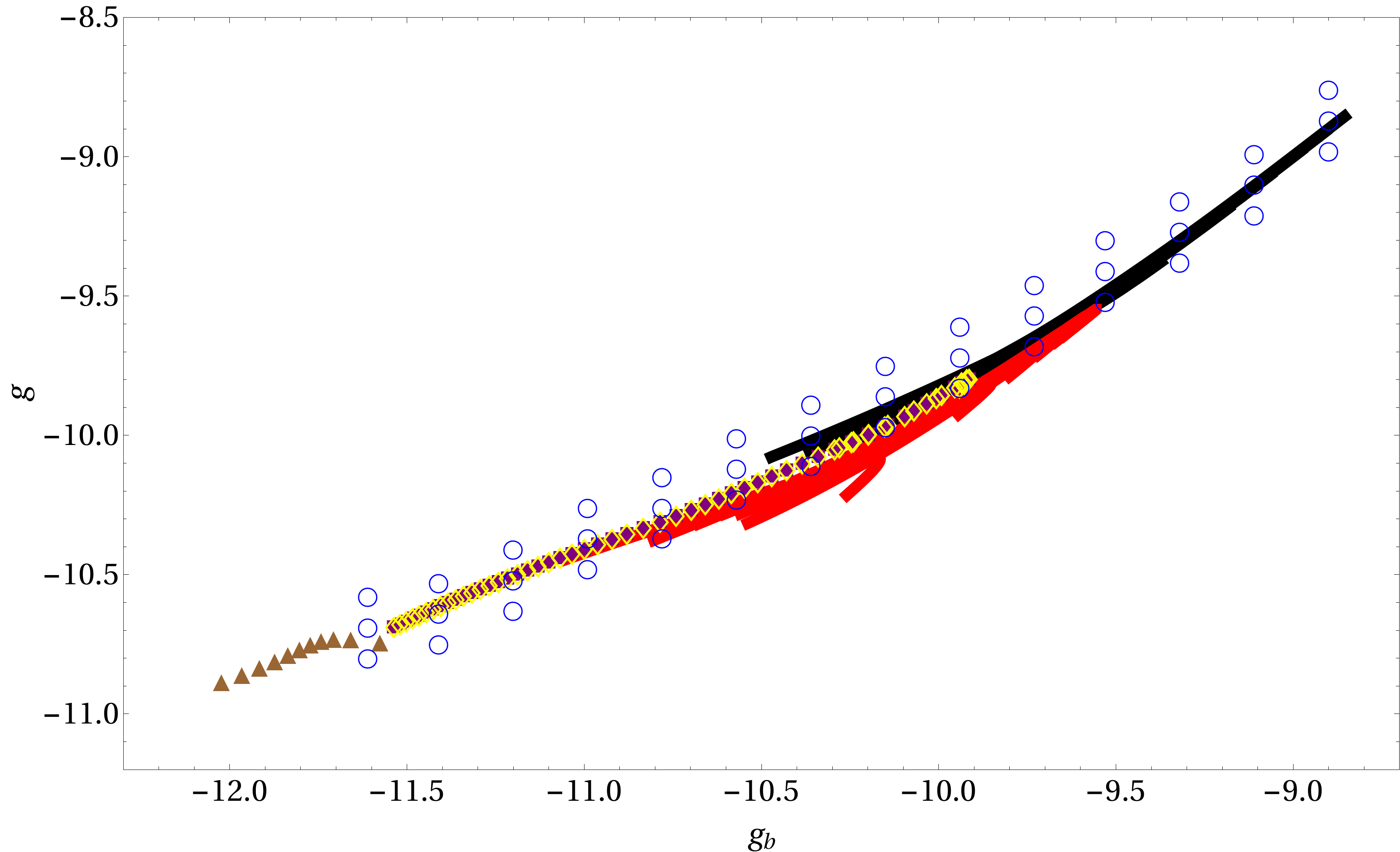}
\end{center}
\caption {The McGaugh et al. 2016 relationship with its 1-$\sigma$ uncertainty ({\it blue circles}) and that obtained by the
URC ({lines}). Accelerations are in units of $Log\ m/s^2$}

\end{figure}

$\bullet$ The URC frames the Mc Gaugh et al relationship, emerging in spirals between the total acceleration $g(r)=V^2(r)/r$ at a radius $r$ from the center and its baryonic component: $g_b(r)=V_b^2(r)/r$, into the cored dark halo scenario. In fact, we can derive the quantities $log \ g_{URC}$ and $ log \ g_{URCb}$ from the URC, entirely rooted in such a scenario, and realize that they lie in the region marked by the McGaugh relationship see Fig(6).

These concordant observational evidences are very difficult, if not impossible, to have them explained in the simple Dark Matter
framework in vogue so far. On the other hand, they are likely beacons and maybe portals to the new Physics that seems to lurk behind the
phenomenon called "Dark Matter"

\section*{Acknowledgments}
PS thanks the organizers of the 18 Lomonosov Conference where this work has taken place.


\begin{thebibliography}{99}

 \bibitem{D} Donato, F., Gentile, G., Salucci, P., 2004, MNRAS, 353L, 17 
\bibitem{e} Evoli, C., Salucci, P., Lapi, A., \& Danese, L.\ 2011, ApJ, 743, 45 
\bibitem{freeman} Freeman, K. C., 1970, ApJ 160, 811
\bibitem{McGaugh} McGaugh, S., Lelli, F. et al, Phys. Rev. Lett. 2016, 117, 201101
\bibitem{PS95} Persic, M \& Salucci, P. 1995, ApjS,99,501 
\bibitem{p91} Persic, M \& Salucci, P. 1991, ApJ, 368, 60.
\bibitem {pss}Persic, M., Salucci, P., Stel, F., 1996, MNRAS 281, 27
\bibitem {rubin} Rubin, V.~C., Ford, W.~K., Jr., Thonnard, N., et al \ 1982, ApJ, 261, 439 
\bibitem {SB}Salucci, P., Burkert, A.2000, ApJ, 537,9
\bibitem {s7} Salucci, P.; Lapi, A.; Tonini, C et al. 2007, MNRAS, 378, 41
\bibitem{tonini} Tonini, C. Lapi, A. Shankar, F. et al  2006, ApJ, 638, 13
\bibitem {ys} Yegorova, I, Salucci., P. 2007, MNRAS, 377, 507


 

\end{thebibliography}
\end{document}